\documentclass[
reprint,
amsmath,
amssymb,
aps,
]{revtex4-2}

\usepackage{graphicx}
\usepackage{dcolumn}
\usepackage{bm}
\usepackage[hyperfootnotes=false]{hyperref}
\usepackage{multirow}
\usepackage[makeroom]{cancel}
\usepackage[export]{adjustbox}
\usepackage{xcolor}
\usepackage{braket}

\newcommand{\fig}[1]{Fig.~\ref{#1}}
\newcommand{\eq}[1]{Eq.~\ref{#1}}
\newcommand{\beq}{\begin{equation}}
\newcommand{\eeq}{\end{equation}}

\renewcommand{\ss}{Schr{\"o}dinger's equation}
\newcommand{\ntwo}{\frac{1}{\sqrt{2}}}
\renewcommand{\L}{\ket{\text{L}}}
\newcommand{\R}{\ket{\text{R}}}
\newcommand{\Rp}{\ket{\text{R}^\prime}}
\newcommand{\Rpn}{\ket{\text{R}^\prime_\text{n}}}
\definecolor{Blue}{rgb}{.0,.15,.85}
\renewcommand{\u}{\ket{\mathcolor{Blue}{\text{unbumped}}}}
\newcommand{\C}{\ket{\mathcolor{Blue}{\text{C}}}}
\newcommand{\Cm}{\ket{\mathcolor{Blue}{\text{C}_\text{m}}}}
\definecolor{Red}{rgb}{.78,.1,.12}
\renewcommand{\b}{\ket{\mathcolor{Red}{\text{bumped}}}}
\renewcommand{\H}{\ket{\mathcolor{Red}{\text{H}}}}
\newcommand{\Hn}{\ket{\mathcolor{Red}{\text{H}_\text{n}}}}
\definecolor{Green}{rgb}{0.25,.55,.2}
\newcommand{\upYou}{\ket{\mathcolor{Green}\Uparrow}}
\newcommand{\downYou}{\ket{\mathcolor{Green}\Downarrow}}
\definecolor{Purple}{rgb}{0.3,0,.7}
\newcommand{\downMe}{\ket{\mathcolor{Purple}\Downarrow}}
\newcommand{\upMe}{\ket{\mathcolor{Purple}\Uparrow}}
\newcommand{\upEyesYou}{\ket{\includegraphics[scale=.12,valign=c]{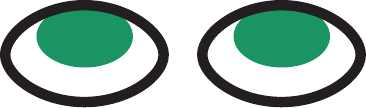}}}
\newcommand{\upEyesYoum}{\ket{\includegraphics[scale=.12,valign=c]{eyesUpYou}_\text{m}}}
\newcommand{\upEyesMen}{\ket{\includegraphics[scale=.12,valign=c]{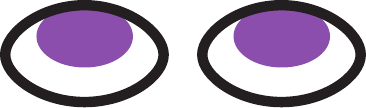}_\text{n}}}
\newcommand{\downEyesYou}{\ket{\includegraphics[scale=.12,valign=c]{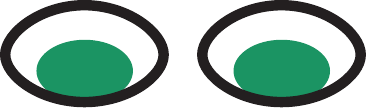}}}
\newcommand{\downEyesYoum}{\ket{\includegraphics[scale=.12,valign=c]{eyesDownYou}_\text{m}}}
\newcommand{\downEyesMen}{\ket{\includegraphics[scale=.12,valign=c]{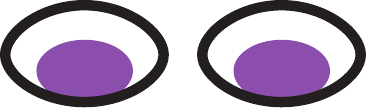}_\text{n}}}

\begin{document}
\preprint{APS/123-QED} 
\title{The Quantum Many-Worlds Interpretation, Simply Told}

\author{Brian C. Odom}
\email{b-odom@northwestern.edu}
\affiliation{Center for Fundamental Physics, Department of Physics and Astronomy, Northwestern University, Evanston, Illinois 60208, USA}

\keywords{quantum, interpretation, everett, many worlds, schrodinger's cat, which-path, bolometer, action at a distance}

\date{\today}

\begin{abstract}
The many-worlds interpretation (MWI) of quantum mechanics poses a simple question.  What would reality look like if everything evolved in time according to the same quantum equations? There is an attractive consistency to treating microscopic objects, measuring devices, and observers all on the same footing, but do the predictions match our observations? Here, we build a model for a bolometer detector making a which-path measurement in an atom interferometer. We discuss the MWI claim that, while both measurement outcomes occur in each experimental iteration, an observer will experience only one outcome or the other with a probability consistent with experiment. Finally, we discuss how MWI does not have action at a distance. This article is written to be accessible to anyone with an undergraduate course in quantum mechanics.
\end{abstract}

\maketitle

\section{Introduction}
Courses in quantum mechanics (QM) typically teach that the wavefunction collapses when a measurement is made. Physicists often refer to this as the `Copenhagen interpretation,' but since historians question the accuracy of that label, we will here call it the `textbook' version of QM. Textbook QM has had great experimental success, and many practitioners see no need for revision. But many physicists find it unsatisfactory. If QM is a good theory, should we not be able to describe the measurement device and observers as part of the quantum system? If so, does collapse ever really occur? On the other hand, if we abandon collapse, are we truly supposed to believe that Schr{\"o}dinger's cat is both alive and dead?

How to understand quantum measurement is still an active area of debate, but some leading alternatives to textbook QM are well established. The many-worlds interpretation (MWI)~\cite{tegmark2007many,carroll2020something,tegmark1998interpretation}, originally put forward in the 1957 Ph.D. thesis of Hugh Everett~\cite{everett1957hugh}, has enough support in the community~\cite{gibney2025physicists} that it is arguably worthy of discussion in QM courses.

MWI can be said to follow from two basic postulates~\footnote{Some variants of MWI have additional postulates.}$^\dagger$. First, \ss\ is obeyed at all times. Second, wavefunctions describe actual particle states, rather than some observer's knowledge of the particles. Said another way, MWI is simply textbook QM as most people think of it, with the measurement postulates removed. In MWI, microscopic objects, measurement devices, and observers are all treated the same. Advocates argue that this consistency is a strong indication that MWI provides a better description of nature, if indeed it can be shown to work.

Another reason to provide an introduction to MWI is that it has become embedded in popular culture, ranging from literature, to movies, to expressions such as ``Maybe in an alternate world..." Popular representations of science often get some things right and others wrong. As current and future representatives of Physics, it is important for QM students to be able to distinguish between accurate portrayals and wrong-headed notions, even if some of the latter do make for good stories.

In this manuscript, we will build a simple model for a which-path detector coupled to an atom interferometer. We shall see that applying deterministic evolution of \ss\ leads to the atom-detector-environment system evolving into an entangled superposition of measurement outcomes. No collapse postulate will be used, but nonetheless an observer perceives the illusion of probabilistic collapse as they become entangled as well. Finally, we will discuss how action at a distance is not a feature of MWI. The question of whether the wavefunction describes physical reality independent of any observer is discussed in App.~\ref{reality}.    

MWI is an active area of research, and not all advocates agree on exactly what the theory is and which problems remain to be solved~\cite{vaidman2024many}. Here, we describe a minimalist version of MWI with broad support in the community. We provide endnotes tagged with a $^\dagger$ at points where opinions diverge. 

\begin{figure*}[htbp!]
    \centering
    \includegraphics[width=0.95\textwidth]{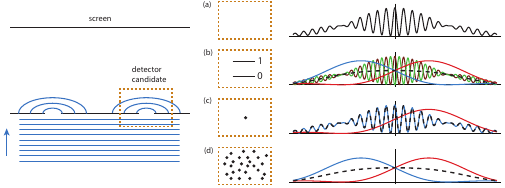}
    \caption{Atom interferometer apparatus with various detector candidates inserted into one arm. The right column shows example atom patterns on the screen, correlated with different detector-candidate states. The screen is placed in the the mid-field regime (instead of the normal far-field) to separate the single-slit patterns from each other. For simplicity, the example patterns correspond to negligible back-action of the detector on the atom.    (a) Nothing inserted. (b) Qubit inserted. Atom patterns: qubit ignored (dashed black), in state $\ket{0}$ (blue), $\ket{1}$ (red), $\ket{+}$ (brown), or $\ket{-}$ (green). (c) Single gas molecule inserted. Atom patterns: molecule ignored (dashed black), in state $\ket{\mathcolor{Blue}{\text{unbumped}}}$ (blue), or $\ket{\mathcolor{Red}{\text{bumped}}}$ (red). (d) Many gas molecules inserted. Atom patterns: molecules ignored (dashed black), in cold state $\C$ (blue), or hot state $\H$ (red).}
    \label{int}
\end{figure*}

\section{Which-Path Detection in an Atom Interferometer}
To see how MWI works, we will develop a quantum description of a which-path detector placed in one arm of an atom interferometer. \textbf{We define a `detector' simply as some coupled system which stores information for later readout.} We expect that which-path detection should always be accompanied by loss of interference fringes, but we shall wait to see what the theory predicts.

We will use a balanced atom interferometer (\fig{int}) described by 
\beq
\label{psi1}
\ket{\psi}=\ntwo\left(\L+\R\right),
\eeq
where $\L$ and $\R$ are the atom states passing through the left and right path. The Born rule says that the probability density at a point $\vec{r}$ on the screen is given by 
\begin{align}
\mathcal{P}&=\left|\braket{\vec{r}|\psi}\right|^2=\frac{1}{2}\left|\braket{\vec{r}|\text{L}}+\braket{\vec{r}|\text{R}}\right|^2 \nonumber\\ 
&=\frac{1}{2}\left|\psi_\text{L}(\vec{r})+\psi_\text{R}(\vec{r})\right|^2 \nonumber \\ 
&=\frac{1}{2}|\psi_\text{L}(\vec{r})|^2+\frac{1}{2}|\psi_\text{R}(\vec{r})|^2+ \textrm{Re}[\psi_\text{L}^*(\vec{r}) \psi_\text{R}(\vec{r})].
\end{align}
In the last expression, the first two terms are the single-path contributions, and the third term is the interference.

\section{A Qubit (Non-)Detector}
First, consider a qubit initialized to $\ket{0}$ and coupled with 100\% efficiency to the right arm (\fig{int}b.) If the atom traverses the right path, the qubit state changes, such that the joint atom-qubit state evolves as 
\beq
\ket{\psi} = \R\ket{0} \rightarrow \Rp\ket{1},
\label{qubitCoupling}
\eeq
where $\ket{\text{R}^\prime}$ differs from $\ket{\text{R}}$ because of back-action of the qubit on the atom. If the atom passes through the left path, the qubit remains unchanged:
\beq
\ket{\psi} = \ket{\text{L}}\ket{0} \rightarrow \ket{\text{L}}\ket{0}. 
\eeq
For the balanced interferometer state of \eq{psi1}, the atom-qubit state evolves into
\beq
\ket{\psi} = \ntwo\left(\L\ket{0}+\Rp\ket{1}\right), 
\label{qubit1}
\eeq
The initially unentangled atom state of \eq{psi1} has become entangled with the qubit. 

Before the interaction, we could discuss the atom and qubit states separately, but we can no longer do so afterwards. Entanglement is sometimes taught as an interesting add-on rather than a central feature of QM, but \textbf{entanglement is central to understanding measurement in MWI.}  

Does the qubit function like a detector? The MWI conception of a measurement is that all outcomes exist in the final entangled wavefunction. At first glance, \eq{qubit1} appears to describe that scenario. The $\ket{0}$ state registers that the atom traveled left, and the $\ket{1}$ state registers that the atom traveled right. Additionally, it seems at first that we have confirmation that a which-path measurement has been performed, because the Born rule says that interference fringes disappear. For example, when the qubit state is $\ket{0}$, using \eq{qubit1}, we have $\mathcal{P}_0=\left|\braket{\vec{r}, 0|\psi}\right|^2\propto\left|\psi_\text{L}(\vec{r})\right|^2$, where the interference term has vanished because $\braket{0|1}=0$. If we ignore the qubit, the atom pattern is proportional to $\mathcal{P}_0+\mathcal{P}_1$, which also has no interference. Encouragingly, this reproduces the textbook QM postulate that the which-path detector need not be read out by a subsequent observer for interference to be destroyed.

However, it is quite natural to think of the qubit in different bases. For instance, we can rewrite \eq{qubit1} as
\beq
\ket{\psi}=\ntwo(\ket{\psi_+}\ket{+}+\ket{\psi_-}\ket{-}),
\eeq
where $\ket{\psi_\pm}=\ntwo\left(\ket{\psi_\text{L}}\pm\ket{\psi_{\text{R}^\prime}}\right)$, and $\ket{\pm}=\ntwo\left(\ket{0}\pm\ket{1}\right)$. 
The atom states $\ket{\psi_\pm}$ describe symmetric and antisymmetric both-path traversal, and $\ket{\pm}$ are the corresponding qubit excitations. Now the qubit appears to be detecting the atom doing something completely different than it did before, namely traveling both paths in one of two manners. Correspondingly, for experimental arrangements where back-action does not spoil things~\cite{scully1991quantum,storey1994path}, the atoms correlated with the qubit in $\ket{+}$ or $\ket{-}$ still exhibit interference fringes, as shown in \fig{int}b. 

This trick of recovering fringes is called making a `quantum eraser,' because the potential to obtain which-path information was destroyed by our readout of the qubit in one particular basis~\cite{scully1991quantum}. For our purposes here, it might be better described as a `quantum revealer,' because the fact that fringe recovery is possible tells us that coupling to a simple object like a qubit can easily hide coherence, but that coherence is never truly lost.

So, the information stored by the qubit is ambiguous. If the qubit is read out one way it provides which-path information, but read out the other way it provides two-path traversal information. We conclude that a coupled qubit could be a useful front-end for a which-path detector, but it is not a which-path detector in itself. Ironically, the qubit fails to be a detector because it is too well behaved. It started in a completely known state, and the experimenter can `rotate' it to perform readout in any basis. In order to get a good detector with an unambiguous measurement basis, we need to inject some ignorance and/or lack of experimental control. 

\section{A Single-Molecule Detector}
Instead of a qubit, let us place a gas molecule in the right path (\fig{int}c.) The molecule is initialized by reaching thermal equilibrium with the walls of a very cold container, so that it is approximately still. The molecule state $\u$ transitions to state $\b$ if struck by the atom, where change of color indicates an energy increase. Both molecule states are well defined for each iteration of the experiment, but we do not know their precise details. And the details are different each iteration. Even so, we can easily imagine distinguishing with high confidence between a molecule that had been bumped and one that had not. 

The atom-molecule system evolves into 
\begin{align}
\ket{\psi} &= \ntwo(\ket{\text{L}}\u + \alpha\Rp\b \nonumber \\
&+\beta\R\u), 
\label{single}
\end{align}
where the third term describes an atom traveling the right path but missing the molecule, and $\alpha \ll \beta$ means that missing is much more likely than colliding. If the bump is significant enough that $\braket{\mathcolor{Blue}{\text{unbumped}}|\mathcolor{Red}{\text{bumped}}}\approx 0$, then when the interaction actually happens, the molecule records information that the atom had traveled the right path. And in this case, the interference fringes again vanish, by identical math as for the qubit read out in the $\ket{0,1}$ basis. But since even a right-path atom most often misses the molecule, there is almost no which-path information encoded in $\u$. Correspondingly, the atom pattern correlated with the molecule in $\u$, as well as correlated with ignoring the molecule altogether, are essentially the same as for the unperturbed interferometer. In other words, this detector is horribly inefficient.

However, the molecule is qualitatively different from just being an inefficient qubit. Like the qubit, we can read out from the molecule which-path information. But in contrast to the qubit, we cannot read out information about symmetric versus antisymmetric two-path traversal. The reason is that it would be impractical to read out a superposition of $\ket{\mathcolor{Blue}{\text{unbumped}}}$ and $\ket{\mathcolor{Red}{\text{bumped}}}$, even if we knew the precise details of those states. And we do not know those precise details. Thanks to our lack of experimental control and our ignorance, the molecule is unambiguously a which-path detector, albeit an inefficient one. 

\section{A Bolometer Detector}
We can create a detector with 100\% efficiency by adding more gas molecules to the volume (\fig{int}d.) We initialize the detector to some cold state, now labeled $\C$, about which we once again do not know microscopic details. This creates a bolometer detector, which encodes which-path information in the energy deposited into the gas. 

It would be nice if we could simply write $\ket{\psi} = \L\C \rightarrow \L\C$ and $\ket{\psi} = \R\C \rightarrow \Rp\H$, where $\H$ represents a hot detector state. Then we would have
\beq
\ket{\psi} = \ntwo\left(\ket{\text{L}}\C+\Rp\H\right).\ \  \color{gray}\text{(naive)}\color{black}
\label{naive}
\eeq
That naive expression has the right spirit of an MWI superposition involving different detector states, but its setup brushed something important under the rug.

What actually happens within the gas is that there are a large number of paths the atom can follow, each leaving the gas in a different hot configuration. The system evolves as
\beq
\ket{\psi} = \R\C \rightarrow \sum_n b_n \Rpn\Hn, \ \  \color{gray}\text{(better)}\color{black}\ 
\eeq
where $\Hn$ describes some gas configuration and $\Rpn$ is the corresponding final atom state. The detector is said to have a complex `internal environment,' with the macroscopic detector state $\H$ corresponding to a great number of microscopically different states $\Hn$. And since each $\Hn$ has a unique $\Rpn$, a right-path atom becomes entangled with this internal environment. Furthermore, both the cold and hot internal environments become entangled with atoms in the container walls, and through those, with the photons, gas, etc. outside the container. So, the detector's `external environment' is also brought into the entanglement. We can write
\begin{align}
&\ket{\psi} = \L\C \rightarrow \sum_m a_\text{m} \L\Cm \ket{\varepsilon_{\text{m}}}  \nonumber\\
&\ket{\psi} = \R\C \rightarrow \sum_n b_\text{n} \Rpn\Hn \ket{\varepsilon_{\text{n}}}, 
\end{align}
where $\ket{\varepsilon_{\text{m,n}}}$ are the external environment states. So, after the measurement, the entire system is in state
\begin{align}
\ket{\psi} &= \ntwo\sum_{\text{m}}a_\text{m}\L\Cm\ket{\varepsilon_{\text{m}}} \nonumber \\ 
&+\ntwo\sum_{\text{n}}b_\text{n}\Rpn\Hn\ket{\varepsilon_{\text{n}}}. \ \  \color{gray}\text{(correct)}\color{black}\
\label{entangled}
\end{align}
The atom has become entangled with the detector's macroscopic cold/hot states, its internal environment, and the external environment.

We now have a model for a 100\% efficient and unambiguous which-path detector. It records the measurement information, encoded in the gas energy, for later readout. Similar to the qubit where $\braket{0|1}=0$, fringe disappearance is a consequence of $\braket{\mathcolor{Blue}{\text{C}_\text{m}}|\mathcolor{Red}{\text{H}_\text{n}}}\approx 0$. But here, for the first time among the detector candidates, the atom interference fringes disappear for each possible detector state or if the detector is ignored. We see that collapse is not necessary for fringes to vanish; this occurs as a natural consequence of entangling with an efficient detector. 

But notice that the state~\eq{entangled} describes a reality where both $\C$ and $\H$ outcomes occur! \textbf{For each single iteration of the experiment, the macroscopic detector evolves into an entangled superposition including all possible measurement outcomes.} This conclusion is not \emph{a priori} assumed by MWI, but is rather an unavoidable consequence of its two postulates.

\section{Decoherence and Branching}
We can verify that simple systems can exist in superposition states by observing interference fringes. If MWI is correct that measurement places macroscopic detectors into superpositions with different outcomes, can this be verified as well? If we could do something like the quantum eraser/revealer trick (from the qubit) to recover interference fringes of an atom which had interacted with the bolometer, that would provide indirect evidence that both detector states had existed in the wavefunction. Alas, this task is impossible because macroscopic detectors, especially considering all their environmental entanglements, are not sufficiently under our experimental control. 

Decoherence theory~\cite{zeh1970interpretation, zurek1991decoherence, schlosshauer2019quantum, odom2025beyond} formalizes our description of how entanglement destroys interference. But the term `decoherence' must not be misunderstood. In MWI, the overall wavefunction always evolves coherently according to \ss. It is only when we are ignorant of details and/or lack control that we experience the illusion that the wavefunction decoheres, i.e. that it loses the ability to interfere.

Decoherence theory also resolves a basis question of MWI. We made an argument, based in part upon intuition of practical laboratory capabilities, that the bolometer has a well-defined measurement basis, unlike the qubit. This basis was comprised of the cold/hot classical states which we used to write \eq{entangled}. But in QM, should we not be able to rewrite \eq{entangled} in any basis we like? If any basis was as good as the other, it would be questionable to claim that the predictions of MWI unambiguously match observations, where detectors are always in classical states. Decoherence theory shows that a single basis emerges as being able to make predictions for massively entangled objects, and this basis is the classical one~\footnote{Some MWI advocates question whether decoherence is the most straightforward way of obtaining the natural basis.}$^\dagger$.  

\eq{entangled} describes a scenario where the cold bolometer states have decohered from the hot ones. In MWI parlance, sets of states which have decohered from one another are described as being on different `branches' of the wavefunction. How many branches are there? That is a matter of taste, since \eq{entangled} describes the actual state, and grouping its terms is only a matter of conceptual convenience. We could say there are two branches describing cold and hot bolometer states. Or (as is more commonly done) we could say that each of those meta-groups contain a huge number of non-interfering branches~\footnote{There exists in the community a number of different ways to define the branches~\cite{vaidman2024many}}$^\dagger$.

\section{Observing the Detector}
\eq{entangled} describes the basic result of MWI, without any need for an observer. The bolometer evolves into a superposition involving all measurement outcomes. There, the essential physics was straightforward scattering theory. But we are interested to see whether MWI correctly predicts the experience of observers, so let us extrapolate to that more complicated physics.

The bolometer described by \eq{entangled} already recorded which-path information, but we now imagine engineering into the apparatus a readout mechanism: a mechanical arrow begins oriented left, and it swings right if the bolometer absorbs energy. We also add to the system a sentient observer, specifically a cat, who is watching the arrow. The microscopic physics of how the arrow swings and how the cat's eyes follow is of course complicated. However, if QM adequately describes either L or R scenario by itself, then for the superposition we have 
\begin{align}
\ket{\psi} &= 
\ntwo\sum_{\text{m}}a_\text{m}\L\Cm\ket{\,\includegraphics[scale=.4,valign=c]{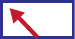}\,_\text{m}}\ket{\,\includegraphics[scale=.1,valign=c]{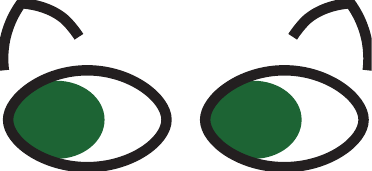}\,_\text{m}}\ket{\varepsilon_{\text{m}}} \nonumber \\
& +\ntwo\sum_{\text{n}}b_\text{n}\Rpn\Hn \ket{\,\includegraphics[scale=.4,valign=c]{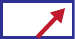}\,_\text{n}}\ket{\,\includegraphics[scale=.1,valign=c]{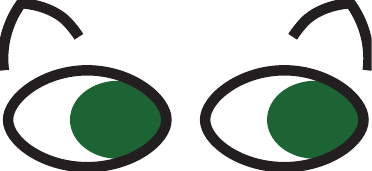}\,_\text{n}}\ket{\varepsilon_{\text{n}}}.
\label{observing}
\end{align}
\textbf{There are now two macroscopically distinct versions of the cat, each experiencing definite (but different) outcomes of the measurement.} 

\section{Collapse Versus World Splitting}
Textbook QM tells us that when we make a measurement, we collapse the wavefunction, and we must set to zero the amplitudes of terms not corresponding to the observed measurement result. MWI disagrees and says we should never set amplitudes to zero by hand. 

Without the detector, reality was described by an interfering superposition of the atom taking two paths. In MWI, inserting a detector results in reality now being described by a non-interfering superposition. In other words, decoherence causes each evolving branch to be forever isolated within its own Hilbert subspace, which is never invaded by another branch. The two groups of terms in \eq{observing} then effectively describe the diverging evolution of two `worlds' which are completely and forever disconnected. One branch tells the story of the atom going left, a cold detector, and the cat looking left. The other branch tells the story of the atom going right, a hot detector, and the cat looking right. 

We can now address a common misconception. Critics sometimes characterize MWI as ``putting in worlds" by hand. \textbf{But in fact, sets of distinct storylines (worlds) naturally emerge whenever uncontrolled entanglements develop.}

MWI advocates enthusiastically acknowledge that the collapse of textbook QM adequately describes the experience of observers. The wavefunction of \eq{observing} did not collapse, but for either version of the cat in its own disconnected world, it might as well have. However, adequacy is no justification for needlessly sacrificing accuracy. 

\section{Whence Probability}
Probabilities are not fundamental to a deterministic theory.  Even so, the use of probabilities is essential for describing the experience of limited observers. This is true of both classical physics and MWI, but what we mean by `limited' differs in interesting ways.

In classical physics, probability arises as an artifact of our ignorance.  If we had perfect knowledge of a system, then we could find the answer to any question about its future by propagating the deterministic equations. But since we often lack sufficient knowledge, we find probabilistic analysis useful, for instance when describing coin tosses or when using statistical mechanics.

Textbook QM claims that nature is intrinsically probabilistic, but MWI disagrees. Like classical physics, MWI is governed by purely deterministic equations. However, unlike in classical physics, observers in MWI with perfect knowledge of the initial state still experience probabilistic outcomes. The reason is that observers themselves become part of the entangled system.

Consider once again the cat observing the bolometer detector, but this time we will make the atom path superposition imbalanced: $\ket{\psi} = \sqrt{\frac{1}{100}}\ket{\text{L}} + \sqrt{\frac{99}{100}}\ket{\text{R}}$. In single-world textbook QM, the measurement has 99\% probability of yielding R. In MWI, the state after the measurement is 
\begin{align}
\label{postM}
\ket{\psi} &= 
\sqrt{\frac{1}{100}}\sum_{\text{m}}a_\text{m}\L\Cm\ket{\,\includegraphics[scale=.4,valign=c]{detL}\,_\text{m}}\ket{\,\includegraphics[scale=.1,valign=c]{catL}\,_\text{m}}\ket{\varepsilon_{\text{m}}} \nonumber \\
& +\sqrt{\frac{99}{100}}\sum_{\text{n}}b_\text{n}\Rpn\Hn \ket{\,\includegraphics[scale=.4,valign=c]{detR}\,_\text{n}}\ket{\,\includegraphics[scale=.1,valign=c]{catR}\,_\text{n}}\ket{\varepsilon_{\text{n}}}.
\end{align}
It would be a mistake to ask before the measurement about the cat's probability to see R. With 100\% probability, one version of the cat will see L, and the other will see R. 

However, there is a time in the process when a probability question arises~\cite{vaidman1998schizophrenic,sebens2018self}. Let us imagine that the cat keeps its eyes closed until a few seconds after the arrow registers an outcome. Before the cat opens its eyes, there are countless microscopic exchanges entangling the cat with the system. For instance, some photons scattered off the left-pointing arrow will strike the cat whereas those same photons will not be scattered from the right-pointing arrow. The eyes-closed cat entangles and splits into different versions $\ket{\,\includegraphics[scale=.1,valign=c]{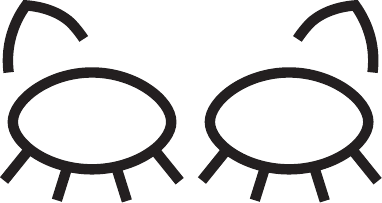}\,^\text{L}}$ and $\ket{\,\includegraphics[scale=.1,valign=c]{catClosed}\,^\text{R}}$. Consciousness of the outcome is not a prerequisite for entanglement and branching. (To exaggerate this point, Vaidman imagines giving the observer a sleeping pill so that they wake an arbitrary time after the detector has registered an outcome~\cite{vaidman1998schizophrenic}. These concepts are closely related to `Wigner's friend' thought experiments~\cite{lawrence2025original}.) The state is
\begin{align}
\ket{\psi} &= 
\sqrt{\frac{1}{100}}\sum_{\text{m}}a_\text{m}\L\Cm\ket{\,\includegraphics[scale=.4,valign=c]{detL}\,_\text{m}}\ket{\,\includegraphics[scale=.1,valign=c]{catClosed}\,^\text{L}_\text{m}}\ket{\varepsilon_{\text{m}}} \nonumber \\
& +\sqrt{\frac{99}{100}}\sum_{\text{n}}b_\text{n}\Rpn\Hn \ket{\,\includegraphics[scale=.4,valign=c]{detR}\,_\text{n}}\ket{\,\includegraphics[scale=.1,valign=c]{catClosed}\,^\text{R}_\text{n}}\ket{\varepsilon_{\text{n}}}.
\end{align}

In this eyes-closed interval, when  an MWI-savvy cat does not yet know which branch it is on, it can ask itself what the likelihood is, that upon opening its eyes, it will see the arrow pointed right. The Born rule says the answer is 99\%. After a few iterations of this experiment, the cat would learn to strongly expect the R outcome, just as in single-world textbook QM. Probabilities \textit{feel} the same whether an observer believes textbook QM or MWI.

However, the status of the Born rule in MWI is still being worked out. In textbook QM, the Born rule can be postulated since collapse is postulated. But for MWI, the most common viewpoint is that no collapse means that the Born rule must be derived~\cite{carroll2020something}. If indeed probabilities are related to amplitudes, the amplitude-squared dependence is easy to prove~\cite{sebens2018self}. But that probabilities and amplitudes are necessarily related is a subtle point, and it is currently a matter of debate whether a satisfactory derivation exists~\cite{vaidman2024many, carroll2020something}~\footnote{Some people think it is alright in MWI to leave the Born rule as a postulate, so that its derivation is not a show-stopping challenge~\cite{vaidman2022many}.}$^\dagger$. 

\section{No Spooky Action at a Distance}
Consider the standard example of a Bell-pair of entangled spins, one sent to \textcolor{Green}{you} and the other sent to \textcolor{Purple}{me}:
\beq
\ket{\psi} = \ntwo \left(\upYou \downMe - \downYou\upMe \right).
\eeq
If I measure my spin first, I have a 50\% chance to find either $\upMe$ or $\downMe$. But in textbook QM, if you measure yours first and observe $\upYou$, then you collapse the two-spin wavefunction. Even if I am a great distance away, your action of measurement has instantly changed my odds to 100\% of measuring $\downMe$. If we interpret the wavefunction as describing the real particle states and not just an observer's knowledge of the particles, your distant measurement has instantly changed the state of my particle~\cite{boughn1806there}. This is what Einstein called `spooky action at a distance.' Since we cannot use entanglement to transmit information faster than the speed of light, there is no technical disagreement with the theory of relativity. But it does seem to violate the spirit of relativity. 

MWI reaches a strikingly different conclusion~\cite{vaidman2015bell}. If you measure your spin first, you will branch into two different versions of yourself $\upEyesYou$ and $\downEyesYou$.
One branch contains $\upYou\upEyesYoum $ terms and the other has $\downYou\downEyesYoum$, with the subscript again representing microscopic configurations. If I subsequently measure my spin, I split and become part of the branching structure you already established. Now one branch contains $\upYou\upEyesYoum\downMe\downEyesMen$ terms and the other has $\downYou\downEyesYoum\upMe\upEyesMen$ terms. Whether or not you measured first does not affect post-splitting-me's 50\% probability of finding either outcome. Since the wavefunction does not collapse in MWI, there is no action at a distance. 


\section{Conclusions}
MWI simply asks what happens if we acknowledge that our detectors and ourselves rightly belong in the wavefunction, right along with everything else. We find that measurements are described by a branching structure in the entangled wavefunction, with different outcomes occurring on different branches. If the wavefunction describes physical reality, rather than some observer's knowledge of reality, then we must conclude that all branches (worlds) are real. 

Branching is irreversible in the same sense as breaking an egg is irreversible in classical physics~\cite{lawrence2022observing}. Both describe a practical limitation of our capabilities, rather than anything fundamental. Irreversibility of branching means that you can never gather evidence that more than one branch actually exists. But what you can do is test smooth evolution of \ss\ on increasingly large systems and look for an unexpected breakdown. Like all good physical theories, MWI is not provable, but it is falsifiable. 

We already discussed how MWI agrees with Einstein's intuition about spooky action at a distance. Regardless of Einstein's preferences, MWI's lack of action at a distance is a mark in its favor. Einstein also famously objected to the randomness of Copenhagen QM, by declaring, ``God does not play dice." Although we do not know what Einstein would have thought of MWI, it once again agrees with him. MWI claims that we need not insert the fundamental randomness of collapse into an otherwise elegant theory, because deterministic evolution of \ss\ predicts all known experimental results. As we observers discover ourselves to be on one branch or the other, we experience the illusion of randomized collapse. God does not play dice, but we do.

\begin{acknowledgments}
We gratefully acknowledge clarifying discussions with Pulak Dutta and Surjeet Rajendran. This work was funded by NSF grant 610-4011500-60062402.
\end{acknowledgments}

\appendix
\section{Equations and Reality}
\label{reality}
The equations of classical physics have undeniable success in making predictions. The position and momentum of an objects obey Newton's differential equation, and electromagnetic fields obey Maxwell's equations. With these tools, we can predict a vast range of physical phenomena.

What does this successes tell us about reality? For classical physics, a `realist' interpretation is usually held implicitly. In this view, Newton's second law is successful because position and momentum are real properties of particles, and Maxwell's equations are successful because the electromagnetic field is a real entity existing at all points in space. Positions and momenta of particles and electromagnetic field values are objective aspects of the universe, independent of any observer's awareness of them.

MWI continues applying this realist thinking to the wavefunction. (Here, we are considering non-relativistic quantum mechanics which replaces Newtonian mechanics, but the same approach is used for interpreting relativistic quantum field theory.) \ss\ is seen as being successful because wavefunctions describe real physical properties of particles. A particle has a well-defined wavefunction, independent of any observer. MWI claims to be the most natural conclusion of wavefunction realists who also think that everything is governed by \ss.

However, the realist interpretation of science is not the only one available. `Epistemic' interpretations argue that solutions to successful equations should not be over-interpreted as mapping onto reality. Consider an experiment which measures the time for an apple to fall from a tree. The expression for the apple's position $x$ that solves $F = m\ddot{x}$ correctly predicts the observed time. An epistemic interpretation would be that, while position is definitely a useful tool for making such predictions, we should not therefore conclude that position is a real property of the apple. This perspective sounds strange to most ears, and indeed epistemic interpretations of classical physics do not have much of a following. But they do enjoy a sizable advocacy in QM. In the QBist epistemic interpretation~\cite{fuchs2000quantum, mermin2012commentary}, a particle's wavefunction is only defined insofar as there is also an associated `agent' (a generalized observer) who will make a measurement. The wavefunction of a particle does not represent the particle state, but rather the agent's knowledge of the particle state. The primary attraction of epistemic interpretations is this: if wavefunction are merely mathematical tools that observers can use to make predictions, then we can avoid the bizarre conclusion of many worlds. 

It is left as an exercise for the reader to find their own conviction on the realist/epistemic question of wavefunction interpretation. But it is worth emphasizing that this is a question upon which our entire interpretation of quantum reality hinges.

\bibliographystyle{aipnum4-2}
\bibliography{bib.bib}
\end{document}